\documentclass[aps,prb,amsmath,amssymb,10pt,superscriptaddress,twocolumn,color,epsfig,graphicx,bm]{revtex4-1}

\usepackage{color}
\usepackage{graphicx}
\usepackage{graphics}
\usepackage{subfigure}
\usepackage{epsfig}
\usepackage{amsmath}
\usepackage{array}
\usepackage{amssymb}
\usepackage{graphicx}      
\usepackage{dcolumn}       
\usepackage{bm}            
\usepackage{array}
\usepackage{booktabs}

\usepackage[linktocpage=true,
  colorlinks=true, 
  pdfborder={0 0 0},
  linkcolor=blue,
  citecolor=red,
  filecolor=yellow,
  urlcolor=blue,
  bookmarks,
  pdfauthor={},
]{hyperref}

\newcommand{\tokyo}{Department of Applied Physics, The University of Tokyo, Tokyo 113-8656, Japan}

\begin{document}

\title{
Periodicity-free unfolding method of electronic energy spectra:\\ Application to twisted bilayer graphene
}

\author{Taichi Kosugi}   \affiliation{\tokyo}
\author{Hirofumi Nishi}   \affiliation{\tokyo}
\author{Yasuyuki Kato}   \affiliation{\tokyo}
\author{Yu-ichiro Matsushita}   \affiliation{\tokyo}


\date{\today}

\begin{abstract}
We propose a novel periodicity-free unfolding method of the electronic energy spectra. Our new method solves a serious problem that calculated electronic band structure strongly depends on the choice of the simulation cell, i.e., primitive-cell or supercell.
The present method projects the electronic states onto the free-electron states, giving rise to the {\it plane-wave unfolded} spectra.
Using the method, the energy spectra can be calculated as a completely independent quantity from the choice of the simulation cell.
We have examined the unfolded energy spectra in detail for three models and clarified the validity of our method: One-dimensional interacting two chain model, monolayer graphene, and twisted bilayer graphene.
Furthermore, we have discussed that our present method is directly related to the experimental ARPES (Angle-Resolved Photo-Emission Spectroscopy) spectra.
\end{abstract}

\pacs{~}
\maketitle

\section{Introduction}
Electronic-structure calculations based on the density functional theory (DFT)
are one of the most powerful tools to elucidate and even predict the electronic properties of condensed matters.
In material science, impurity doping, alloying, and surface deposition and adsorption are widely used for manipulating the electronic properties of target materials.
In particular, impurity doping is of high technological importance.
In order to treat such imperfectness in a crystal,
DFT calculations often adopt the supercell approach
and it is known to be efficient for comprehensive understandings of material properties.
In that approach, one has to increase the size of the unit cell to mimic the randomness and/or to simulate the low density of impurities within periodic boundary conditions.

However, when we use the supercell scheme, it causes a serious problem.
A supercell is by definition larger than the primitive cell
and hence the corresponding Brillouin zone (BZ) is smaller than the primitive-cell BZ.
The larger the chosen supercell is for faithful simulation of a perturbed system,
the denser the calculated band structure are in reciprocal space,
hindering the direct comparison between angle-resolved photo-emission spectroscopy (ARPES) experiments\cite{ARPES_Suga_Sekiyama} and band calculations.
As a prescription for solving such a dilemma,
the band unfolding method was proposed 
~\cite{PhysRevLett.104.216401, PhysRevLett.104.236403,PhysRevB.71.115215} and has been widely used particularly in DFT calculations.
By virtue of this conventional band unfolding technique,
we can compare the calculated band structure with experiments and reasonable agreements have been reported \cite{1367-2630-17-8-083010}.
This method enables one to analyze the various perturbative effects on a periodic system such as
disorder and defects\cite{0953-8984-26-15-155502, PhysRevB.90.115202, PhysRevB.92.125146}.
A more generic formulation of the unfolding applicable to two-component wave functions\cite{PhysRevB.91.041116}
and those from the group theoretical viewpoint\cite{1367-2630-16-3-033034, PhysRevB.95.024305}
have been proposed.
The spirit of the conventional unfolding approach is also applicable to phonon spectra\cite{PhysRevB.90.205214, 0953-8984-27-30-305402, ZHENG2016218}.

However, as we have already pointed out\cite{unfolding_bilayer_graphene_submitted}, the conventional unfolding method has a big issue to provide unphysical/artificial ghost bands in the unfolded bands in applying it to multi-periodicity materials, such as piled thin films and two-dimensional materials on a substrate. In them, there exist multiple possible primitive-cell BZs. The modified unfolding method reported in Ref.~\onlinecite{unfolding_bilayer_graphene_submitted} was demonstrated to correct dramatically the wrong behaviors inherent in the conventional unfolding method and to give physically reasonable results for such materials. 

Still the unfolding schemes possess a serious problem in its practical applications that the unfolded energy spectra strongly depend on the choice of the primitive cell to which the electronic bands are unfolded.
In addition, a much severe calculation target might be
alloys, because they don't have a clear primitive cell anymore. 
In such cases, we cannot apply the periodicity-assumed  unfolding schemes. The energy band structure should be determined independently from how we prepare the simulation cell. To satisfy the physical condition, the unfolding methods defined without any assumption of the primitive cell (periodicity-free) are strongly desired. 
In present study, we propose a plane-wave unfolding method, which does not assume the existence of a primitive cell at all. The periodicity-free method allows us to draw even the electronic band structures which are essentially aperiodic. The aim of this study is to propose the novel band unfolding method and to demonstrate successful results for three simple models using tight-binding (TB) calculations: One-dimensional interacting two chain model, monolayer graphene, and twisted bilayer graphene.
We have to mention, however, that our new formulation can be used also in DFT calculations without any change. 

This paper is organized as follows. In Section II, we derive the expressions for the unfolded spectral function in linear combination of atomic orbitals (LCAO) picture employed throughout the present study. The relation between the new method with ARPES simulations are also discussed. In Section III, we apply the new method to three systems which are two chains, monolayer graphene, and tBLG using TB calculations. The various kinds of contributions to the spectral functions are intensively analyzed to understand the features of the calculated spectra. In Section IV, the conclusions are provided.

\section{Methods}
\subsection{Expressions}

We begin with the projected spectral function defined by
\begin{gather}
	A (\lambda, \varepsilon)
	\equiv
		-\frac{1}{\pi}
		\mathrm{Im}
		\mathrm{Tr}
		\left[ \hat{\mathcal{P}}_{\lambda} \frac{1}{\varepsilon + i \eta - \hat{H}} \right]
	,
    \label{def_gen_spec}
\end{gather}
where the trace is taken over the electronic states for the target system, whose one-body Hamiltonian is $\hat{H}$.
$\eta$ is an infinitesimal positive constant.
$\hat{\mathcal{P}}_\lambda$ is the projection operator onto the Hilbert space spanned by the states characterized by continuous parameter(s) $\lambda$ in reciprocal space
chosen so that the Hilbert space is expected to be helpful for the analyses.
For a case where the target system is a perturbed but periodic system
and $\lambda$ is a crystal momentum $\boldsymbol{k}$ for the unperturbed system,
the spectral function defined above reduces to the original unfolded spectra proposed by Ku et al.\cite{PhysRevLett.104.216401}
For a case where the target system is aperiodic and $\lambda$ designates a vector in reciprocal space together with the localization in real space\cite{unfolding_bilayer_graphene_submitted},
our definition is also suitable.
These observations indicate that our spectral function assumes neither periodicity nor perturbed nature in target systems.
Therefore we refer to it as the generalized unfolded spectral function.

The conventional unfolding techniques employ the supercell method in which a target system is assumed to be a weakly perturbed or disordered but is essentially periodic.
While such systems ensure the successful application of the existing unfolding method as useful tools for the analyses of large periodic systems,
those consisting of subsystems having incommensurate periodicities hinder the straightforward applications of the methods to them.
In this study,
we focus on the unfolding method with $\lambda$ being the wave vector of a plane wave
since one of our purposes is to demonstrate that the plane-wave unfolding is useful also for a system consisting of subsystems with truly or nearly incommensurate periodicities. 
The spectral function projected onto a free-electron state having a wave vector $\boldsymbol{k}_{\mathrm{f}}$ for a periodic target system is given by
\begin{gather}
	A (\boldsymbol{k}_{\mathrm{f}}, \varepsilon)
	=
		\sum_{\boldsymbol{k}, m}
			| \langle \boldsymbol{k}_{\mathrm{f}} | \psi_{\boldsymbol{k} m} \rangle |^2 \delta ( \varepsilon_{\boldsymbol{k} m} - \varepsilon )
	\label{def_spec_pw}
\end{gather}
as a special case for eq.~(\ref{def_gen_spec}) using $\hat{\mathcal{P}}_{\boldsymbol{k}_{\mathrm{f}}} = | \boldsymbol{k}_{\mathrm{f}} \rangle \langle \boldsymbol{k}_{\mathrm{f}} |$.
$| \psi_{\boldsymbol{k} m} \rangle$ is the $m$-th energy eigenstate belonging to the eigenvalue $\varepsilon_{\boldsymbol{k} m}$ for a crystal momentum $\boldsymbol{k}$ defined within the FBZ.
We ignore the spin degree of freedom in the present study for simplicity.
Since the free-electron states span the complete orthonormalized set for one-electron states, that is,
$\sum_{\boldsymbol{k}_{\mathrm{f}}} | \boldsymbol{k}_{\mathrm{f}} \rangle \langle \boldsymbol{k}_{\mathrm{f}} | = \hat{1}$,
the unfolded spectral function integrated over the wave vectors is equal to the electronic density of states.
This means that all the original information on the energy spectra obtained in an electronic-structure calculation is conveyed to the unfolded spectral function without any loss.
It can be equivalently said that the original band structure in a periodic system is reconstructed by folding the plane-wave unfolded spectra.
We have therefore adopted the terminology "plane-wave unfolding" associating with the previous unfolding techniques.

If one tries to perform electronic-structure calculations of an incommensurate-periodicity material,
a periodic system consisting of sufficiently large unit cells is practically often prepared for mimicking the incommensurability.
The prescription described below is thus needed also for such calculations.

We assume that the electronic state of the three-dimensional periodic target system is described accurately in LCAO picture.
The generic expressions for the conventional unfolding in LCAO picture have been provided\cite{0953-8984-25-34-345501}.
The method described below can be applicable to low-dimensional systems with slight modifications.
The Bloch sum of the $\mu$-th localized basis $| \phi_{\boldsymbol{R} \mu} \rangle$ within a unit cell located at a lattice point $\boldsymbol{R}$ is
$
| \phi_{\boldsymbol{k} \mu} \rangle
=
N_{\mathrm{cells}}^{-1/2}
\sum_{\boldsymbol{R}}
e^{i \boldsymbol{k} \cdot \boldsymbol{R}}
| \phi_{\boldsymbol{R} \mu} \rangle
,
$
where $N_{\mathrm{cells}}$ is the number of unit cells contained in the system.
The energy eigenstate is expanded in the Bloch sums using the eigenvector $\boldsymbol{c}_{\boldsymbol{k} m}$ of the one-body Hamiltonian as
$
| \psi_{\boldsymbol{k} m} \rangle
=
\sum_\mu
c_{\boldsymbol{k} \mu m}
| \phi_{\boldsymbol{k} \mu} \rangle
.
$
Substitution of this expression into eq.~(\ref{def_spec_pw}) leads to
\begin{gather}
	A ( \boldsymbol{k}_{\mathrm{f}}, \varepsilon)
	\propto
		\sum_m
			\int_{\mathrm{FBZ}}
			d^3 k \,
	\cdot
	\nonumber \\
	\cdot
				\Bigg|
					\sum_\mu
					c_{\boldsymbol{k} \mu m}
					e^{-i \boldsymbol{k}_{\mathrm{f}} \cdot \boldsymbol{\tau}_\mu }
					\widetilde{\phi}_\mu (\boldsymbol{k}_{\mathrm{f}})
					\sum_{\boldsymbol{R}}
					e^{i (\boldsymbol{k} -  \boldsymbol{k}_{\mathrm{f}} ) \cdot \boldsymbol{R}}
				\Bigg|^2
				\delta (\varepsilon - \varepsilon_{\boldsymbol{k} m} )		
	,
	\label{spec_sum_over_integers}
\end{gather}
where $\boldsymbol{\tau}_\mu$ is the relative position of the site within a unit cell
and the integration is performed over the first BZ (FBZ).
$
\widetilde{\phi}_\mu (\boldsymbol{k}_{\mathrm{f}})
\equiv
\int
d^3 r \,
	e^{-i \boldsymbol{k}_{\mathrm{f}} \cdot \boldsymbol{r} }
	\phi_{\mu} (\boldsymbol{r})
$
is the Fourier component of the basis function localized at the origin.
Using the primitive lattice vectors $\boldsymbol{a}_1, \boldsymbol{a}_2,$ and $\boldsymbol{a}_3$,
the formula of the delta functions 
$
\sum_{n = -\infty}^\infty
\delta (x - n)
=
\sum_{n = -\infty}^\infty
e^{i 2 \pi n x}
$
for an arbitrary $x$,
the summation over lattice points in the equation above is calculated as
\begin{gather}
	\sum_{\boldsymbol{R}}
		e^{i (\boldsymbol{k} -  \boldsymbol{k}_{\mathrm{f}} ) \cdot \boldsymbol{R}}
	=
		\prod_j
		\sum_{n_j = -\infty}^\infty
			\delta \left( \frac{(\boldsymbol{k} -  \boldsymbol{k}_{\mathrm{f}} ) \cdot \boldsymbol{a}_j}{2 \pi} - n_j \right)	
	.
\end{gather}
Since there exists only a single combination of $n_j (\boldsymbol{k}_{\mathrm{f}})$'s and
$\boldsymbol{k}_{(\mathrm{f})}$ in the FBZ for an arbitrary $\boldsymbol{k}_{\mathrm{f}}$ such that
$
(\boldsymbol{k}_{(\mathrm{f})} -  \boldsymbol{k}_{\mathrm{f}} ) \cdot \boldsymbol{a}_j - 2 \pi n_j (\boldsymbol{k}_{\mathrm{f}}) = 0 
$
for all the $j$'s,
the $\boldsymbol{k}$ integration in eq.~(\ref{spec_sum_over_integers}) can be performed to give
\begin{gather}
	A ( \boldsymbol{k}_{\mathrm{f}}, \varepsilon)
	\propto
		\sum_m
		I_m (\boldsymbol{k}_{\mathrm{f}})
		\delta (\varepsilon - \varepsilon_{\boldsymbol{k}_{(\mathrm{f})} m} )	
	,
\end{gather}
where
\begin{gather}
	I_m (\boldsymbol{k}_{\mathrm{f}})
	=
		\sum_{\mu, \mu'}
		\gamma_{\mu \mu'}^{\mathrm{atom}} (\boldsymbol{k}_{\mathrm{f}})
		\gamma_{\mu \mu'}^{\mathrm{geom}} (\boldsymbol{k}_{\mathrm{f}})
		\gamma_{m \mu \mu'}^{\mathrm{Bloch}} (\boldsymbol{k}_{(\mathrm{f})})
	\label{generic_pbc_intensity_from_band}
\end{gather}
is the spectral intensity coming from the $m$-th band.
We have factorized the intensity into
\begin{gather}
	\gamma_{\mu \mu'}^{\mathrm{atom}} (\boldsymbol{k}_{\mathrm{f}})
	\equiv
			\widetilde{\phi}_\mu (\boldsymbol{k}_{\mathrm{f}})
			\widetilde{\phi}_{\mu'} (\boldsymbol{k}_{\mathrm{f}})^*
	\label{generic_pbc_def_gamma_atom}
\end{gather}
depending only on the shapes of the basis functions,
\begin{gather}
	\gamma_{\mu \mu'}^{\mathrm{geom}} (\boldsymbol{k}_{\mathrm{f}})
	\equiv
		e^{-i \boldsymbol{k}_{\mathrm{f}} \cdot ( \boldsymbol{\tau}_\mu - \boldsymbol{\tau}_{\mu'}) }
	\label{generic_pbc_def_gamma_geom}
\end{gather}
depending only on the relative positions of the atoms,
and
\begin{gather}
	\gamma_{m \mu \mu'}^{\mathrm{Bloch}} (\boldsymbol{k}_{(\mathrm{f})})
	\equiv
			c_{\boldsymbol{k}_{(\mathrm{f})} \mu m}
			c_{\boldsymbol{k}_{(\mathrm{f})} \mu' m}^*
	,
	\label{generic_pbc_def_gamma_Bloch}
\end{gather}
which in contrast depends on the electronic structure of the periodic system.
This factorization tells us that the damped behavior of an unfolded spectra with an increasing $k_{\mathrm{f}}$ comes only from the shapes of the localized orbitals,
regardless of the crystal structure.

For a simple case in which the target system consists of repeated primitive cells each of which contains only a single localized orbital,
the geometric and the Bloch parts of the intensity in eq.~(\ref{generic_pbc_def_gamma_Bloch})  are identically unity:
$\gamma^{\mathrm{geom}} (\boldsymbol{k}_{\mathrm{f}}) = 1$ and 
$\gamma^{\mathrm{Bloch}} (\boldsymbol{k}_{(\mathrm{f})}) = 1$,
indicating that the intensity is determined only by the atomic part.
In particular, if the localized orbital is approximated as an $s$-type Gaussian function, whose the Fourier component is always decreasing and nonzero,
the intensity is ensured to be nonzero on an iso-energy plot of the spectral function.

Approximation of all the localized orbitals in the target system as having the same shape
allows one to factor out the decaying term and to define
\begin{gather}
	I_m^{\mathrm{cryst}} (\boldsymbol{k}_{\mathrm{f}})
	\equiv
		\sum_{\mu, \mu'}
		\gamma_{\mu \mu'}^{\mathrm{geom}} (\boldsymbol{k}_{\mathrm{f}})
		\gamma_{m \mu \mu'}^{\mathrm{Bloch}} (\boldsymbol{k}_{(\mathrm{f})})
\end{gather}
instead of the true intensity in Eq.~(\ref{generic_pbc_intensity_from_band}).
This function is useful for simple analyses for capturing the features of the spectral function
since it does not decay and its non-periodicity in reciprocal space can come only from $\gamma^{\mathrm{geom}}$.
For a system in which this approximation is good, we are lead to the following two insights.
First, the presence or absence of the periodicity of $I_m^{\mathrm{cryst}}$ in reciprocal space is determined only by the relative positions of the atoms.
Second,  if multiple atoms exist in the primitive cell and the periodicity of $I_m^{\mathrm{cryst}}$ exists,
the periodicity is larger than the primitive-cell BZ since every vector connecting the atoms in the unit cell is inside it [see Eq.~(\ref{generic_pbc_def_gamma_geom})].      
The direct consequence of the second insight is found in monolayer graphene, as demonstrated later.

If the localized basis functions are the so-called Cartesian Gaussian functions\cite{Helgaker},
the expression for the atomic contribution to the spectral intensity can further be factorized.
We provide the explicit expressions in Appendix.

\subsection{Relation with ARPES simulations}

We discuss here the relation between the plane-wave unfolding and ARPES simulations.
The theoretical studies for describing a photoemission process began from the three-step model\cite{PhysRev.136.A1044},
in which the process consists of the photoelectron excitation, the photoelectron transport to the surface, and the photoelectron escape out of the sample.
Later the one-step model\cite{HOPKINSON198069, PENDRY1976679}, where the process is "compressed" to a single step, was proposed
and has been continuing to be modified as well as the three-step model to incorporate the correlation effects and/or the relativistic effects (see, e.g., Ref.~\onlinecite{0953-8984-16-26-026}).

Puschnig and L\"uftner\cite{PUSCHNIG2015193} performed simulations of ARPES images from the results of  DFT calculations using plane-wave basis set
by adopting the one-step model and assuming the final state to be a plane wave\cite{PhysRevB.10.5030}.
Specifically, they used the following expression for the ARPES spectra:
\begin{gather}
	I (\boldsymbol{k}_{\mathrm{f}}, \omega)
	=
		\sum_{\boldsymbol{k}, m}^{\mathrm{occ.}}
		|\boldsymbol{A} \cdot \boldsymbol{k}_{\mathrm{f}} |^2
		| \langle \boldsymbol{k}_{\mathrm{f}} | \psi_{m \boldsymbol{k} }\rangle |^2
		\delta (\varepsilon_{\boldsymbol{k} m} + \Phi + E_{\mathrm{kin}} - \omega)
	,
	\label{intensity_ARPES_one_step_pw}
\end{gather}
where $\boldsymbol{k}_{\mathrm{f}}$ is the momentum of the photoelectron in the final state having the kinetic energy $E_{\mathrm{kin}} = k_{\mathrm{f}}^2/(2 m)$,
$\boldsymbol{A}$ is the polarization vector of an incident photon having a frequency $\omega$,
and $\Phi$ is the work function.
The contributions coming from the factors involving the polarization vector in eq. (\ref{intensity_ARPES_one_step_pw}) are called the matrix elements effects.
Their simulated spectra for graphene exhibited good agreement with the experiments.

Moser\cite{Moser201729} recently demonstrated that ARPES spectra can be well reproduced for
various systems including monolayer graphene within TB calculations.
He derived the expressions for the spectral function by taking into account the matrix elements effects and the surface states from which the photoelectrons jump into the detector.
The final state was assumed to be well approximated as a plane wave as well as by Puschnig and L\"uftner\cite{PUSCHNIG2015193}.
The central part of his expressions is the momentum distribution of the localized orbitals,
which is mathematically equivalent to the plane-wave unfolding,
and his simulated spectra for graphene look quite similar to those obtained in the present study. 

The ARPES intensity within the one-step model using the plane-wave final state given by eq. (\ref{intensity_ARPES_one_step_pw}) with the matrix element effects removed
essentially coincides with the plane-wave unfolded intensity. 
Although we introduced the plane-wave unfolding originally as a tool for the analyses of computational results and it has nothing to do with any physical process,
this coincidence suggests that the plane-wave unfolding can also be a useful tool for comparison with ARPES experiments.
We should, however, keep in mind that
the assumption that the final state of the photoelectron can be approximated accurately as a plane-wave has often been criticized\cite{PERMIEN1983527,RICHARDSON1983390,1367-2630-17-1-013033}
by stating that it is oversimplification.
If it is the case for a target system, we need to resort to more rigorous methods incorporating the nonequilibrium nature of the photoemission processes\cite{1402-4896-2004-T109-005} such as time-dependent DFT\cite{doi:10.1021/acs.jctc.6b00897}.

\section{Applications}
\subsection{Two chains}

\begin{figure}[htbp]
\begin{center}
\includegraphics[keepaspectratio,width=7cm]{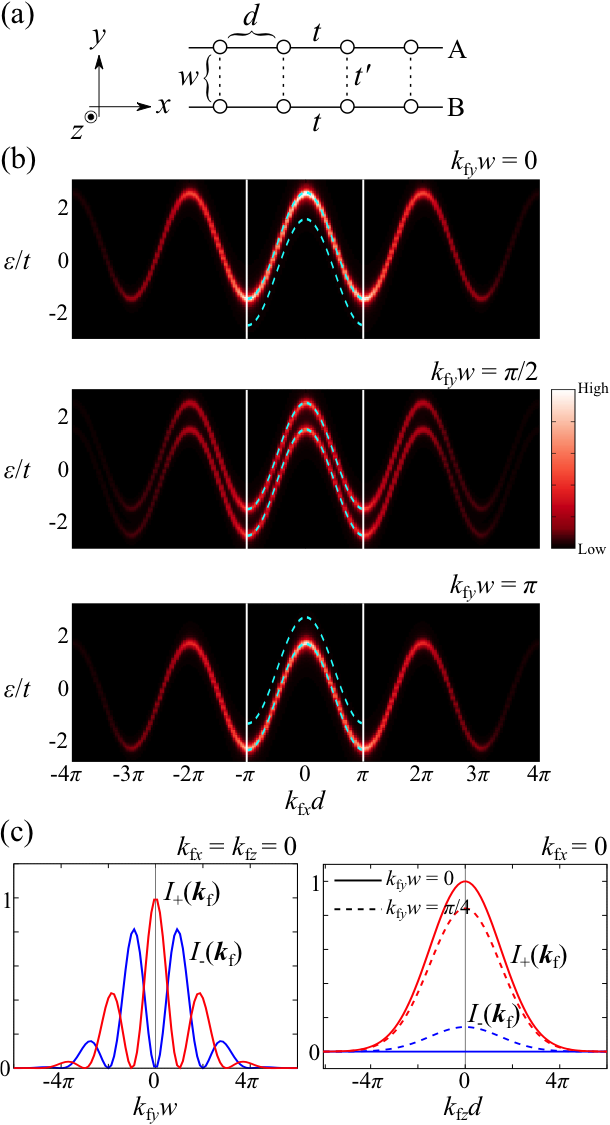}
\end{center}
\caption{
(a) Two chains A and B having infinite lengths along the $x$ axis.
$d (w)$ is the intrachain (interchain) intersite distance. 
(b)
The plane-wave unfolded spectral function $A(\boldsymbol{k}_{\mathrm{f}}, \varepsilon)$ with $k_{\mathrm{f} z} = 0$ for
$k_{\mathrm{f} y} w = 0, \pi/2$, and $\pi$.
The region containing the origin surrounded by the vertical lines represents the FBZ,
inside which the band structure is drawn as dashed curves. 
$t'/t = 0.5, w/d = 1$, and $\sigma/d = 0.15$ were used.
(c)
The spectral intensities $I_\pm (\boldsymbol{k}_{\mathrm{f}})$ as functions of the wave vector of a plane wave.
}
\label{Fig1}
\end{figure}

To capture the characteristics of our new method,
we examine here a simple TB model for infinite-lengths interacting two chains [see Fig.~\ref{Fig1} (a)].
Since this periodic system is analytically solvable,
it helps one to understand clearly the differences between the ordinary band structure and the unfolded spectra onto plane-wave states.

The unit cell consists of two sites at each of which an $s$-type orbital is localized.
We assume that the localized orbitals do not overlap with each other.
$t (t')$ is the real intrachain (interchain) transfer integral between the nearest-neighboring sites.
The geometric part of the spectral intensity in eq.~(\ref{generic_pbc_def_gamma_geom}) is calculated as
$
\gamma_{\mathrm{A A}}^{\mathrm{geom}} (\boldsymbol{k}_{\mathrm{f}})
=
\gamma_{\mathrm{B B}}^{\mathrm{geom}} (\boldsymbol{k}_{\mathrm{f}})
=
1
,
\gamma_{\mathrm{A B}}^{\mathrm{geom}} (\boldsymbol{k}_{\mathrm{f}})
=
\gamma_{\mathrm{B A}}^{\mathrm{geom}} (\boldsymbol{k}_{\mathrm{f}})^*
=
\exp (-i k_{\mathrm{f} y} w)
.
$

The Hamiltonian matrix components for a one-dimensional crystal momentum $k_x$ are
$
H_{\mathrm{A A}} (k_x) = H_{\mathrm{B B}} (k_x) = 2 t \cos (k_x d)
,
H_{\mathrm{A B}} (k_x) = H_{\mathrm{B A}} (k_x) = t'
,
$
whose eigenvalues are $ \varepsilon_{k_x}^{\pm} = 2 t \cos (k_x d) \pm t'.$
The eigenstates are thus given by
$
| k_x, \pm \rangle
=
N
\sum_{m = -\infty}^\infty
e^{i k_x m d} 
( | m, \mathrm{A} \rangle \pm | m, \mathrm{B} \rangle )
,
$
where $| m, j \rangle$ represents the orbital localized at the $m$-th lattice point on  the chain $j$.
$N$ is the normalization constant.
The Bloch part of the spectral intensity in eq.~(\ref{generic_pbc_def_gamma_Bloch}) is thus calculated as
$\gamma_{+ j j'}^{\mathrm{Bloch}} (\boldsymbol{k}_{(\mathrm{f})}) = 1$
for
$j, j' = \mathrm{A}, \mathrm{B}$
and
$
\gamma_{- \mathrm{A A}}^{\mathrm{Bloch}} (\boldsymbol{k}_{(\mathrm{f})}) 
=
\gamma_{- \mathrm{B B}}^{\mathrm{Bloch}} (\boldsymbol{k}_{(\mathrm{f})}) 
=
1,
\gamma_{- \mathrm{A B}}^{\mathrm{Bloch}} (\boldsymbol{k}_{(\mathrm{f})}) 
=
\gamma_{- \mathrm{B A}}^{\mathrm{Bloch}} (\boldsymbol{k}_{(\mathrm{f})}) 
=
-1
.
$

We adopt a normalized $s$-type Gaussian function
$\phi (\boldsymbol{r}) = (\pi \sigma^2)^{-3/4} \exp [-r^2/(2 \sigma^2)]$
with a width $\sigma$ as the basis function.
The atomic part of the spectral intensity in eq.~(\ref{generic_pbc_def_gamma_atom}) is thus calculated as
$
\gamma_{j j'}^{\mathrm{atom}} (\boldsymbol{k}_{\mathrm{f}})
=
2 ( \pi \sigma^2)^{-1/2}
\exp ( -k_{\mathrm{f}}^2 \sigma^2 )
$
for
$j, j' = \mathrm{A}, \mathrm{B}$.
The intensities coming from the branches in eq.~(\ref{generic_pbc_intensity_from_band}) are then, ignoring the common factors,  given by
\begin{gather}
	I_+ (\boldsymbol{k}_{\mathrm{f}})
	=
		e^{-k_{\mathrm{f}}^2 \sigma^2}
		\cos^2 \frac{k_{\mathrm{f} y} w}{2}
	,
	I_- (\boldsymbol{k}_{\mathrm{f}})
	=
		e^{-k_{\mathrm{f}}^2 \sigma^2}
		\sin^2 \frac{k_{\mathrm{f} y} w}{2}
	.
\end{gather}

The unfolded spectral functions with some fixed $k_{\mathrm{f} y}$'s
are shown in Fig.~\ref{Fig1} (b) as a function of $k_{\mathrm{f} x}$.
Their isotropic damped behavior for an increasing magnitude of $\boldsymbol{k}_{\mathrm{f}}$ comes only from the shape of the localized basis function.
It is seen that the intensities for the two branches exhibit anti-phase oscillations as a function of $k_{\mathrm{f} y}$ [see the left panel in Fig.~\ref{Fig1} (c)]. 
This effect is attributed to the interference between the sublattices and
essentially the same effect occurs also in bilayer graphene, as will be shown below.
The intensities as functions of $k_{\mathrm{f} z}$ in contrast exhibit a monotonous decrease [see the right panel in Fig.~\ref{Fig1} (c)]
since there is no sublattice in the $z$ direction.
\subsection{Monolayer graphene}
\begin{figure*}[htbp]
\begin{center}
\includegraphics[width=0.8\linewidth]{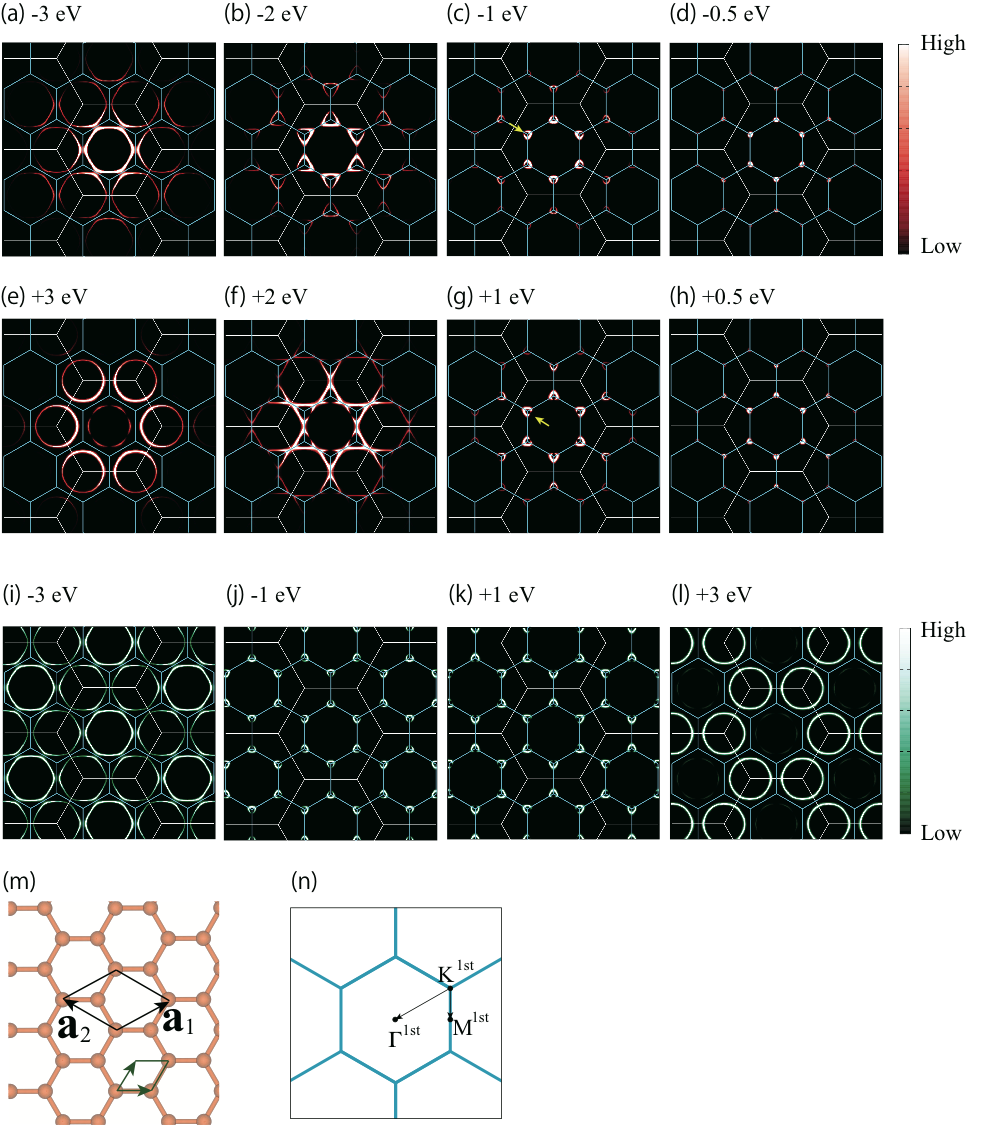}
\end{center}
\caption{
(a)-(h)
The plane-wave unfolded spectral function $A (\boldsymbol{k}_{\mathrm{f}}, \varepsilon)$ with $k_{\mathrm{f} z} = 1.8$ \AA${}^{-1}$
for various $\varepsilon$'s.
The blue hexagons represent the primitive-cell BZs,
while the white ones represent the sublattice cell BZs.
(i)-(l)
The spectral function without the atomic contribution $\gamma^{\mathrm{atom}} (\boldsymbol{k}_{\mathrm{f}})$.
(m)
The conventional primitive cell (black lines)
and
the sublattice cell (green lines).
(n)
The BZs.
That containing $\Gamma^{\mathrm{1st}}$ is the first BZ.
}
\label{Fig2}
\end{figure*}

Let us consider the monolayer graphene as an archetype of the two-dimensional systems with a sublattice structure. We truncate the transfer integrals \cite{PhysRevB.86.125413} up to the eighth-nearest neighbor hopping for simplicity. The monolayer graphene is well described by the TB model on a honeycomb lattice, in which the conventional primitive cell [black lines in Fig.~\ref{Fig2} (m)] has two sublattices. In this paper, we consider the $\pi$-orbitals only; the localized wave function around each site is approximated as a $p_{z}$-type orbital, $\phi (\boldsymbol{r}) = N_\sigma \exp [ -r^2/ (2 \sigma^2) ] z/r$, where $N_\sigma$ is the normalization constant. We set $\sigma = 0.44 l$ with the carbon-carbon distance $l$ as a realistic parameter from a DFT calculation.

We have calculated the unfolded energy spectra for the monolayer graphene. As an overall feature of them, the spectral intensity is getting smaller with increasing the distance from the $\Gamma$ point of the 1st BZ. Another notable point is the existence of the Dirac cones at K points, as is well known in the monolayer graphene. The existence of the Dirac cones causes many intriguing electronic properties such as anomalous quantum Hall effect\cite{Hall_eff1,Hall_eff2}. 
Figure \ref{Fig2} (a)-(h) shows the spectral function $A(\boldsymbol{k}_{\mathrm{f}}, \varepsilon)$ on the $(k_{\mathrm{f} x}, k_{\mathrm{f} y})$ plane by changing $\varepsilon$, for which $k_{\mathrm{f} z}$ is fixed at $1.8$ \AA$^{-1}$. If the $k_{\mathrm{f} z}$ is set to be zero, the energy spectra should be zero due to the symmetry of the $\pi$ orbitals, as discussed below. We find high-intensity circles around K points of the primitive-cell BZ (depicted by the apexes of the blue-colored hexagons) near the Fermi energy, which indicate the conical intersections of the Dirac cones on the iso-energy surface.

The most interesting point here is that the high-intensity circles near the Fermi energy have deficits (incomplete circles) as pointed by an yellow arrow in Fig.~\ref{Fig2} (c) and (g), for example. We found that this incompleteness of the circler spectra originates from a phase factor due to the two sublattice structure. Since the Hamiltonian matrix for each crystal momentum can be diagonalized analytically, the spectral intensities coming from the two branches are calculated, by considering only the nearest-neighboring transfers and ignoring the common factors, as
\begin{gather}
	I_\pm^{\mathrm{NN}} ( \boldsymbol{k}_{\mathrm{f}} )
	=
		\gamma^{\rm atom} (\boldsymbol{k}_{\mathrm{f}})
		\left|
			\mp
			\frac{\rho (\boldsymbol{k}_{(\mathrm{f})})}{|\rho (\boldsymbol{k}_{(\mathrm{f})}) | }
			e^{i \boldsymbol{k}_{\mathrm{f}} \cdot ( \boldsymbol{\tau}_\mathrm{A} - \boldsymbol{\tau}_\mathrm{B} )}
			+
			1
		\right|^2
		,
	\label{monolayer_intensity}
\end{gather}
where $\rho (\boldsymbol{k}_{(\mathrm{f})}) \equiv 1 + \exp [ {i \boldsymbol{k}_{(\mathrm{f})} \cdot \boldsymbol{a}_1} ] + \exp [ i \boldsymbol{k}_{(\mathrm{f})} \cdot \boldsymbol{a}_2 ]$
for the primitive lattice vectors
$\boldsymbol{a}_1 = (3 l/2, \sqrt{3}l/2,0)$ and
$\boldsymbol{a}_2 = (-3 l/2, \sqrt{3} l/2,0)$.
$\boldsymbol{\tau}_\mathrm{A} = (0, 0, 0)$ and $\boldsymbol{\tau}_\mathrm{B} = (-l/2, \sqrt{3} l/2, 0)$ are the positions of the carbon atoms forming the two sublattices. The deficit of energy spectra around a given $\boldsymbol{k}_{\mathrm{f}}$ is the direct consequence of the second factor in the right-hand side of Eq.~(\ref{monolayer_intensity}).
This feature is expected to be common in systems having sublattices. 

Figure \ref{Fig2} (i)-(l) also present the calculated energy spectra without the decay factor, $\gamma^{\rm atom} (\boldsymbol{k}_{\mathrm{f}})$. Looking at the unfolded bands carefully, one can see that the unfolded spectra do not exhibit the periodicity of the primitive-cell BZ (represented by the blue hexagons), but another larger periodicity depicted by the white colored hexagons.
We have found that these white hexagons correspond to the BZ of a sublattice-cell as shown by the green arrows in Fig.~\ref{Fig2} (m).
When we move from a point $\boldsymbol{k}_{\mathrm{f}}$ in reciprocal space by $\Delta \boldsymbol{k}_{\mathrm{f}} = \Delta_1 \boldsymbol{b}_1 + \Delta_2 \boldsymbol{b}_2$ expressed in the reciprocal primitive lattice vectors $\boldsymbol{b}_1$ and $\boldsymbol{b}_2$,
the phase factor in eq.~(\ref{monolayer_intensity}) changes by $\Delta \boldsymbol{k}_{\mathrm{f}} \cdot (\boldsymbol{\tau}_{\mathrm{A}} - \boldsymbol{\tau}_{\mathrm{B}}) = 2 \pi (\Delta_1 + 2 \Delta_2)/3$.
This expression clearly tells us that the spectral intensity is invariant only when $\Delta \boldsymbol{k}_{\mathrm{f}}$ is a reciprocal lattice vector with $\Delta_1 + 2 \Delta_2$ being a multiple of 3,
which is the reason for the larger periodicity of the $\gamma^{\rm atom}$-ignored intensity in reciprocal space than the primitive-cell BZ.
These spectral features, missing spectra and super periodicity in the reciprocal space, are already reported by ARPES measurements \cite{PhysRevB.77.195403, Bostwick2007, PhysRevLett.107.166803,0953-8984-26-33-335501} and the reasons for this deficit were discussed as the matrix element effects.
However, our calculations clearly manifest that the circle-opening is attributed to the symmetry coming from the existence of another primitive cell and is inherent in the electronic bands itself of the monolayer graphene. 
It is recently demonstrated that the missing spectra can be reproduced
using the conventional unfolding by adopting the sublattice cell\cite{Ozaki_unfolding_graphene_submitted},
which gives fair agreement with our descriptions above.

Another noteworthy point is that the position of the missing point is the opposite side between above and below the Fermi energy: Below the Fermi energy, the deficit point is located on the segment K-$\Gamma$ line, whereas above the Fermi energy it is in the opposite direction. This behavior is also observed in experiments\cite{PhysRevB.77.195403, Bostwick2007, PhysRevLett.107.166803,0953-8984-26-33-335501}. 

\begin{figure*}[htbp]
\begin{center}
\includegraphics[keepaspectratio,width=16cm]{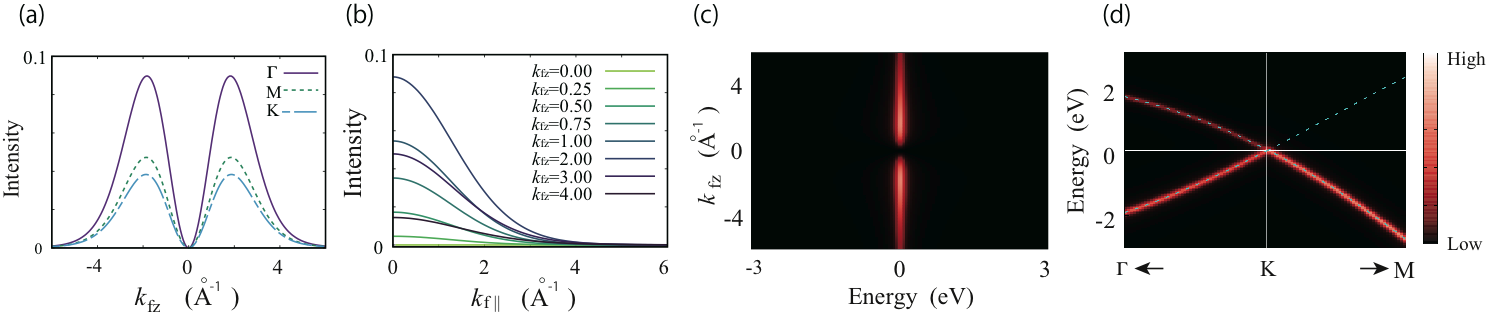}
\end{center}
\caption{
(a), (b)
$\gamma^{\mathrm{atom}} (\boldsymbol{k}_{\mathrm{f}})$ in and out of the ($k_{\mathrm{f} x}, k_{\mathrm{f} y}$) plane.
(c)
The spectral function at the K point with varying $k_{\mathrm{f} z}$.
(d)
The spectral function for fixed $k_{\mathrm{f} z} = 1.8$ \AA$^{-1}$ along the path specified in Fig.~\ref{Fig2} (n). Blue dashed lines represents conventional energy bands of the monolayer graphene.
}
\label{Fig3}
\end{figure*}

In Fig.~\ref{Fig3} (a) and (b), we show $\gamma^{\mathrm{atom}} (\boldsymbol{k}_{\mathrm{f}})$ in and out of the ($k_{\mathrm{f} x}, k_{\mathrm{f} y}$) plane, respectively.
By using the damped oscillatory function
\begin{gather}
	\mathcal{F} (\zeta)
	\equiv
		\frac{1}{\zeta^2}
		\left[
			\frac{\zeta}{\sqrt{2}}
			-
			(\zeta^2 + 1)
			\int_0^{\zeta/\sqrt{2}}
			d t \,
				e^{t^2}
		\right]
	,
\end{gather}
it is expressed as $\gamma^{\mathrm{atom}} (\boldsymbol{k}_{\mathrm{f}}) = |\widetilde{\phi} (\boldsymbol{k}_{\mathrm{f}})|^2 \propto \mathcal{F} (k_{\mathrm{f} \sigma})^2 (k_{\mathrm{f} z} / k_{\mathrm{f}})^2$.
For fixed $k_{\mathrm{f} x}$ and $k_{\mathrm{f} y}$ in figure (a), the intensity takes the maximum value around $k_{\mathrm{f} z}=1.8$ \AA$^{-1}$.
For a fixed $k_{\mathrm{f} z}$ in figure (b), the damping factor is monotonically decreasing as increasing the magnitude of $\boldsymbol{k}_{\mathrm{f}}$ vector. We have also plotted the $k_{\mathrm{f} z}$-dependence of the spectral function at the K point in Fig.~\ref{Fig3} (c).
As clearly seen, at the $k_{\mathrm{f} z}=0$ the energy spectra have no intensity at all.
Corresponding to figure (a), the maximum value achieves around $k_{\mathrm{f} z}= 1.8$ \AA$^{-1}$.
We have plotted the spectral function along the path specified in Fig.~\ref{Fig2} (n), as shown in Fig.~\ref{Fig3} (d).
In the figure, the clear linear dispersion of a Dirac cone appears at the K point.
However, corresponding to the missing spectra mentioned above, one branch is missing above the Fermi energy.

\subsection{Twisted bilayer graphene}
In this subsection, we consider a tBLG where two sheets of graphene are stacked with a twist angle $\theta$ via the van der Waals interaction. The size of the unit cell for tBLG can be arbitrarily changed by tuning $\theta$, and the size can be even infinite to form incommensurate tBLGs. In this study, we set $\theta=9.43^\circ$, so that the number of sites in the unit cell is 148. We set the interlayer distance to $d = 3.349$ \AA \cite{PhysRevB.86.125413, PhysRevB.90.155451}. Consequently, an enormous number of bands are folded in a tiny supercell BZ. However, in practice, it is rarely easy to analyze such bands because they are crossing each other in a very complex way (a typical example is demonstrated in Ref.~\onlinecite{nishi2016}). As explained in Ref.~\onlinecite{unfolding_bilayer_graphene_submitted}, tBLG is one of the systems that the conventional band unfolding method is not applicable to, because the system has multi periodicities. Two primitive-cell BZs corresponding to each monolayer exist in the system and that the conventional unfolding methods do not work properly. We, therefore, applied the new method to tBLG to disentangle such a difficult situation. 


\begin{figure*}[htbp]
\begin{center}
\includegraphics[width=0.8\linewidth]{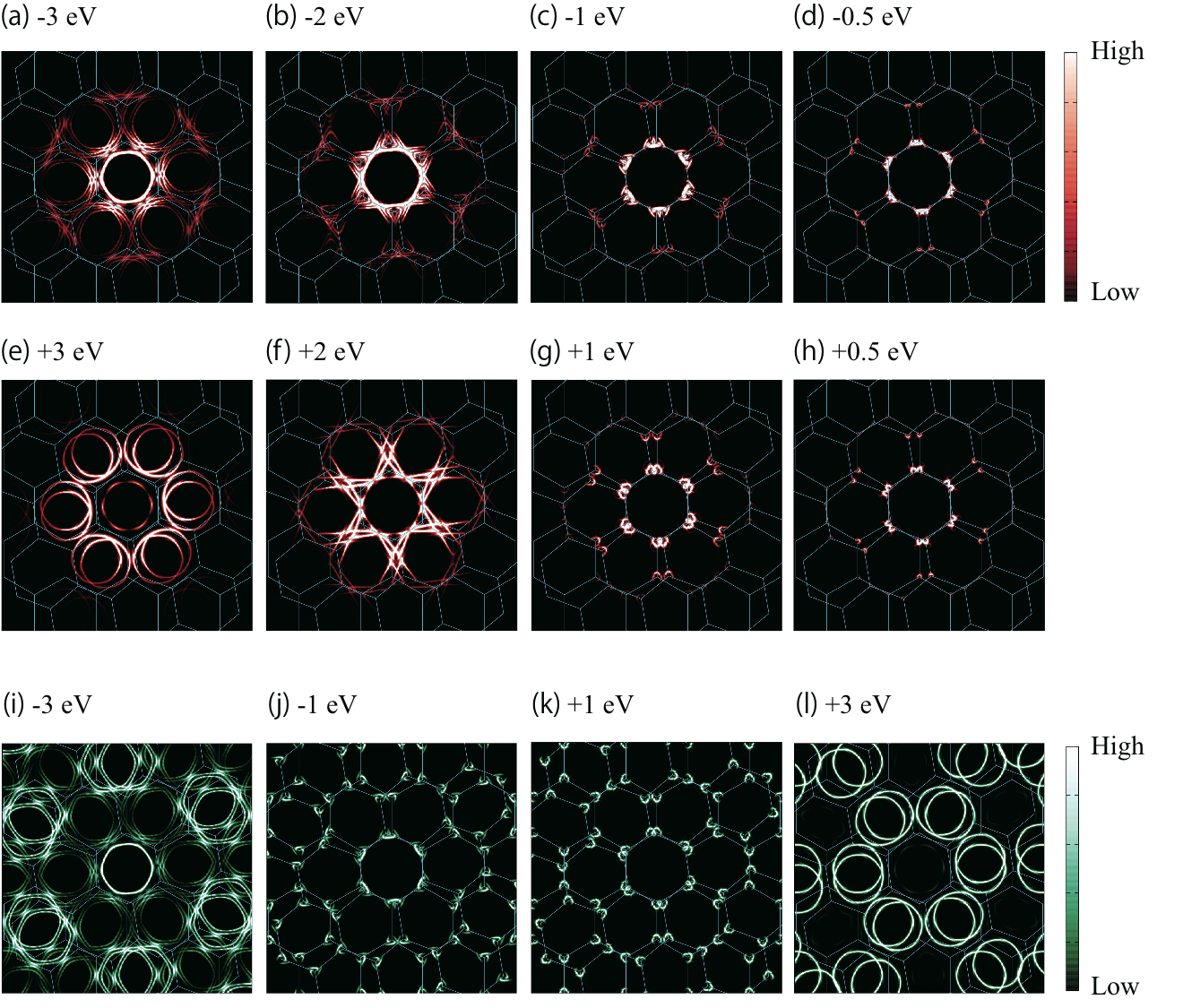}
\end{center}
\caption{
(a)-(h) The plane-wave unfolded spectral function $A(\boldsymbol{k}_{\mathrm{f}}, \varepsilon)$ with $k_{\mathrm{f} z}= 2 \pi/d$ for various $\varepsilon$'s.
The blue hexagons represent the primitive-cell BZs of the two monolayer graphenes.
(i)-(l) The spectral function without the atomic contribution $\gamma^{\mathrm{atom}}(\boldsymbol{k}_{\mathrm{f}})$.
}
\label{Fig4}
\end{figure*}

Fig.~\ref{Fig4} shows the unfolded energy spectra of the tBLG. As one of the most important features, the unfolded spectra show Dirac cones at the K points of the primitive-cell BZs of each monolayer graphene. The unfolded Dirac cones show a similar behavior as those of monolayer graphene: The unfolded spectra also exhibit incomplete circles and circular opening positions are opposite below and above the Fermi energy. The overall feature can be understood as the superposition of the energy spectra of the independent two monolayer graphenes. The unfolded spectra do not exhibit the periodicity of the conventional primitive-cell BZ but that of each sublattice unit cell. Fig.~\ref{Fig4}(i)-(l) show the unfolded spectra without the damping factor $\gamma^{\mathrm{atom}}$. As clearly seen, the distance between adjacent two Dirac points is getting far from each other with increasing the magnitude of $\boldsymbol{k}_{\mathrm{f}}$. 

\begin{figure*}[htbp]
\begin{center}
\includegraphics[keepaspectratio,width=12cm]{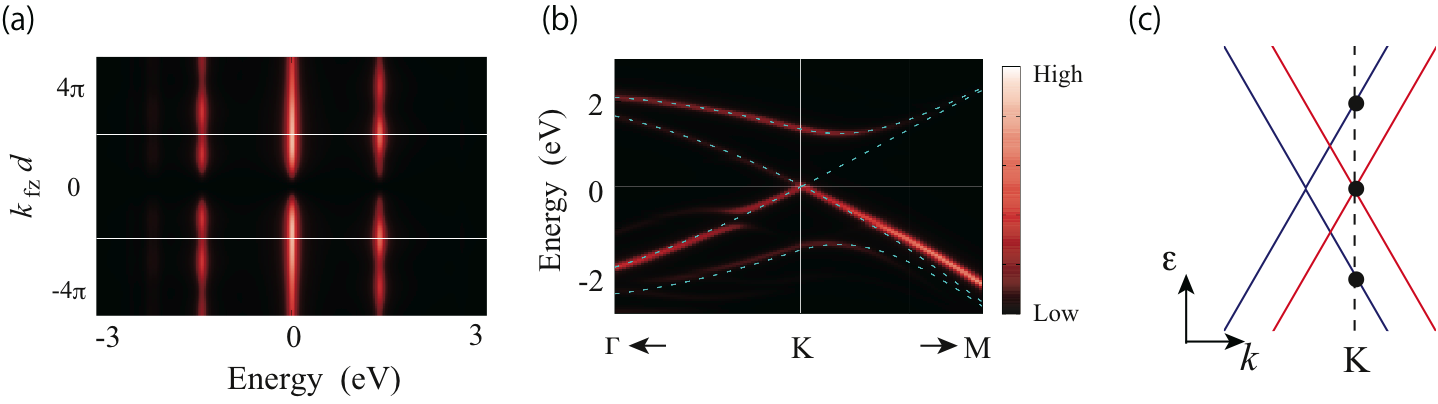}
\end{center}
\caption{
(a) The spectral function for various $k_{\mathrm{f} z}$'s at the K point, with $d$ being the interlayer distance. (b) The spectral function at $k_{\mathrm{f} z}= 2 \pi/d$ along the same high-symmetry path specified in Fig.~\ref{Fig2} (n). (c) Schematic picture of the two adjacent Dirac cones in tBLG. Four electronic bands across the K point (the dashed line) at three intersections. Blue dashed lines represent energy bands of the tBLG without interlayer interaction.
}
\label{Fig5}
\end{figure*}

Fig.~\ref{Fig5} (a) shows the spectral intensity along the $k_z$ direction around a K point.
As fig.~\ref{Fig5} (c) shows a schematic picture in which two adjacent Dirac cones across the K point,
three intersections exist at the K point with different eigen energies.
The three bands clearly appear in figure (a).
The band at $\varepsilon=0$ represents the Dirac point at the K point,
while the other two bands are derived from the branches of the other adjacent Dirac cone.
Interestingly, the intensity of the latter two electronic bands is oscillating with damping.
Such $k_z$ dependence of the two-dimensional electronic bands has been ignored so far. Fig.~\ref{Fig5} (b) shows the unfolded energy dispersion along the same $k$-path in Fig.~\ref{Fig2} (n).
The result shows energy splitting at certain $k$ points.
This result shows good agreement with the previous works \cite{nishi2016, unfolding_bilayer_graphene_submitted},
which gives the validity of our new unfolding method even for multi-periodicity materials. 

\section{Conclusions}
In conclusion, we propose a novel periodicity-free unfolding method of the electronic energy spectra. In principle, the energy band structure should be determined independently from whether we prepare the simulation cell by the primitive cell or by the supercell.
Our new method does satisfy the physical condition. The present method projects the electronic states onto the free-electron states, giving rise to the {\it plane-wave unfolded} spectra.
We derived the expressions of the unfolded spectral function in LCAO picture employed in practical calculations.
It was demonstrated that the spectral intensity is factorized into three contributions, which are the atomic, the geometric, and the Bloch parts.
We examined the plane-wave unfolding by applying it to the TB models for the two chains, the monolayer graphene, and the tBLG.
The unfolded spectra for the two chains, despite the simplicity of the model, was found to exhibit the typical behavior of plane-wave unfolded spectra, that is the oscillation of the intensity coming from the interferences between the sublattices.

For monolayer graphene, we analyzed the unfolded spectra by adopting the $p_z$-type orbital as the basis function and derived the expression for the intensity.
We found that the missing spectra around on the iso-energy surfaces are formed due to the sublattices, consistent with the earlier reproductions in ARPES simulations.
We demonstrated that the larger periodicity in reciprocal space than the primitive-cell BZ originates from the specific relative positions of the sublattices.

Next, we have checked the validity of our plane-wave unfolding method for tBLG, to which the conventional method is not applicable due to its multi-periodicity nature.
Our method successfully produced the unfolded electronic bands of tBLG and unveiled that the electronic bands have also missing spectra inherent in each constituting monolayer graphene.

The application of the new method to DFT calculations is straightforward. The new method applied to electronic-structure calculations for various systems having aperiodic nature such as
defects
will allow us to understand the experiments more clearly than the conventional method.
In particular, the plane-wave unfolding is suitable for systems with defects since the spectral intensity calculated in the conventional method should vanish in the low-density limit.
The new method will be useful also for analyses of variations in the electronic structure of a periodic system with changing its cell parameters since the changes in the shape of BZ complicate the direct comparison between the ordinary band structures.

\section*{Acknowledgments}
This research was supported by MEXT as Exploratory Challenge on Post-K computer (Frontiers of Basic Science: Challenging the Limits).
Y. M. acknowledges the support (partly) from JSPS Grant-in-Aid for Young Scientists (B) Grant Number 16K18075.

\appendix

\section{Spectral intensity for Gaussian-type basis functions}

If one work with a TB calculation, our plane-wave unfolding method requires the explicit expressions for the localized basis functions in real space.
Although what shapes are assumed for the unfolding is arbitrary in principle,
we adopt here the Cartesian Gaussian function of the form
\begin{gather}
	\phi_\mu^{\mathrm{CG}} (\boldsymbol{r})
	=
		N_\mu
		x^{n_{\mu x}}
		y^{n_{\mu y}}
		z^{n_{\mu z}}
		e^{- \frac{r^2}{2 \sigma_\mu^2}}
	,
\end{gather}
where $N_\mu$ is the normalization constant for this Gaussian-type function with its width $\sigma$.
This function is used for represent a localized orbital having an orbital angular momentum $l_\mu \equiv n_{\mu x} + n_{\mu y} + n_{\mu z}$.
Since the basis functions in this form are often adopted in fields of quantum chemistry\cite{Helgaker},
they are expected to give us reliable insights into the unfolded spectra from TB calculations.

By using the definition of the Hermite polynomial
$H_n (x) \equiv (-1)^n 	e^{x^2} ( d/d x)^n e^{-x^2}$,
we can calculate an integral
\begin{gather}
	\int_{-\infty}^\infty
	d x \,
		x^n
		\exp \left(  -\frac{x^2}{2 \sigma^2} - i k x \right)
	\nonumber \\
	=
		\left( i \frac{\partial}{\partial k} \right)^n
		\int_{-\infty}^\infty
		d x \,
			\exp \left(  -\frac{x^2}{2 \sigma^2} - i k x \right)
	\nonumber \\
	=
		\sqrt{2 \pi \sigma^2}
		\left( -i \frac{\sigma}{\sqrt{2}} \right)^n
		e^{-k^2 \sigma^2/2}
		H_n \left( \frac{k \sigma}{\sqrt{2}} \right)
\end{gather}
for a non-negative integer $n$.
With this, the Fourier components $\widetilde{\phi}_\mu^{\mathrm{CG}} (\boldsymbol{k}_{\mathrm{f}})$ of the basis functions can be calculated analytically
and the atomic contribution to the intensity in eq. (\ref{generic_pbc_def_gamma_atom}) is factorized as
\begin{gather}
	\gamma_{\mu \mu'}^{\mathrm{atom}} (\boldsymbol{k}_{\mathrm{f}})
	=
		N_{\mu \mu'}
		\gamma_{\mu \mu'}^{\mathrm{atom-damp}} (k_{\mathrm{f}})
		\gamma_{\mu \mu'}^{\mathrm{atom-osc}} (\boldsymbol{k}_{\mathrm{f}})
	,
\end{gather}
where
\begin{gather}
	\gamma_{\mu \mu'}^{\mathrm{atom-damp}} (k_{\mathrm{f}})
	\equiv
		e^{-k_{\mathrm{f}}^2 (\sigma_\mu^2 + \sigma_{\mu'}^2)/2}	
\end{gather}
and
\begin{gather}
	\gamma_{\mu \mu'}^{\mathrm{atom-osc}} (\boldsymbol{k}_{\mathrm{f}})
	\equiv
		(-i)^{l_{\mu}}
		i^{l_{\mu'}}
		\prod_{\nu = \mu, \mu'}
		\prod_{j = x, y, z}
		H_{n_{\nu j}} \left( \frac{k_{\mathrm{f} j} \sigma_\nu}{\sqrt{2}} \right)
\end{gather}
have been defined.
$N_{\mu \mu'}$ is a constant independent of $\boldsymbol{k}_{\mathrm{f}}$.
$\gamma^{\mathrm{atom-damp}}$ is the only factor responsible for the isotropic damped behavior of the spectral intensity in reciprocal space,
while $\gamma^{\mathrm{atom-osc}}$ is the only factor responsible for the oscillatory behavior coming from the anisotropic shapes of the basis functions.

\bibliographystyle{apsrev4-1}
\bibliography{paper}

\onecolumngrid

\end{document}